\documentclass[preprint]{elsarticle}

\usepackage{amssymb}
\usepackage{amsthm}

\usepackage{lineno}

\journal{Physica B}

\begin{document}

\begin{frontmatter}

\title{Phonon induced optical gain in a current carrying two-level quantum dot}

\author[sbu,ipm]{Amir Eskandari-asl\corref{cor1}}
\ead{amir.eskandari.asl@gmail.com}
\address[sbu]{Department of physics, Shahid Beheshti University, G. C. Evin, Tehran 1983963113, Iran}
\address[ipm]{School of Nano Science, Institute for Research in Fundamental Sciences (IPM),P.O.Box: 19395-5531, Tehran, Iran}

\begin{abstract}
In this work we consider a current carrying two level quantum dot(QD) that is coupled to a single mode phonon bath. Using self-consistent Hartree-Fock approximation, we obtain the I-V curve of QD. By considering the linear response of our system to an incoming classical light, we see that depending on the parametric regime , the system could have weak or strong light absorption or may even show lasing. This lasing occurs at high enough bias voltages and is explained by a population inversion considering side bands, while the total electron population in the higher level is less than the lower one. The frequency at which we have the most significant lasing depends on the level spacing and phonon frequency and not on the electron-phonon coupling strength.
\end{abstract}


\end{frontmatter}

%\linenumbers

\section{Introduction}
One of the most interesting issues in applied physics is the light matter interaction. Recent advances in nano technology made it possible to investigate the interaction of nano-meter sized objects with light. Quantum dot systems have been studied both experimentally\cite{weber2010probing, frey2012dipole, basset2013single, liu2014photon, liu2015semiconductor, stockklauser2015microwave} and theoretically\cite{jin2011lasing, kulkarni2014cavity, gullans2015phonon, karlewski2016lasing} and it has been shown that these systems could be used as light sources or amplifiers. The electron-phonon interaction is shown to assist the optical gain.

In these theoretical studies, a multi-level system is considered that is connected to electronic sources and drains and couples to a continues phonon bath and cavity photon fields. They usually use equations of motion in the Lindblad form to consider the effect of all environments on the system. However, if we want to manipulate molecular sized QDs, it is the case to consider a single mode phonon bath, as it is done in investigating molecular junctions\cite{galperin2008nuclear}. Moreover, we could apply the methods of non-equilibrium Green's functions(NEGF)\cite{Stefan} to make it possible to study the effect of bias voltage more accurately.
 
In this work, we theoretically consider a two-level QD connected to two leads and coupled to a single mode phonon bath. Using Hartree-Fock(HF) approximation we obtain the I-V curve of our system for different electron coupling strength. At a fixed bias voltage, by investigating linear response of the QD to an incoming classical light, we show that depending on the coupling strength and applied bias voltage, we could have weak or strong light absorption and at high enough bias voltages the system shows lasing at some special frequencies. Even though for the optical gain we don't have total population inversion and the higher level is less populated than the lower one, some kind of population inversion occurs between the side-bands, which explains the lasing in our model.  

The paper is organized as follow. In sec.1 we introduce our Hamiltonian and describe the HF approximation. In sec.2 we consider linear response to a classical electric field and obtain the formula for one photon absorption(OPA) cross section. In sec.3 we present our numerical results and sec.4 concludes our work.

\section{Hamiltonian model and the method}
  Our system consists of a QD with two orbitals with different parities, connected to two leads. Each orbital is spin degenerate and coupled to a single mode phonon bath. The Hamiltonian of the model in the presence of external electromagnetic field is
\begin{eqnarray}
\hat{H}_{T}=\hat{H}^{0}_{dot}+\hat{H}_{leads}+\hat{H}_{tun}+\hat{H}^{0}_{ph}+\hat{H}_{e-ph}+\hat{H}_{e-l}.
\label{eq:h1} 
\end{eqnarray}
The first two terms are, respectively, the effective Hamiltonian of a non-interacting two-orbital QD and isolated leads, given by
\begin{eqnarray}
\hat{H}^{0}_{dot}=\sum_{i, \sigma}\epsilon_{i}\hat{c}_{i\sigma}^{\dag} \hat{c}_{i\sigma}
\label{eq:h2}
\end{eqnarray}   
and
\begin{eqnarray}
\hat{H}_{leads}=\sum_{k\in\lbrace R,L\rbrace, \sigma}\epsilon_{k}\hat{a}_{k\sigma}^{\dag} \hat{a}_{k\sigma},
\label{eq:h3}
\end{eqnarray}  
where $ \sigma $ is the spin index, $ \hat{c}_{i\sigma} $($ \hat{c}_{i\sigma}^{\dag}  $) and $ \hat{a}_{k\sigma} $($ \hat{a}_{k\sigma}^{\dag}  $) are, respectively, the annihilation (creation) operators of electron with spin $ \sigma $ of the i-th orbital of QD and k-th level of leads and $ \epsilon_{i} $ is the energy of i-th orbital.

The third term in Eq.\ref{eq:h1} describes  the hybridization between the orbitals of QD and the two leads. For spin independent hybridization it has the form
\begin{eqnarray}
\hat{H}_{tun}=\sum_{ k\in\lbrace R,L\rbrace , i,\sigma }(-t_{ik} \hat{a}_{k\sigma}^{\dag} \hat{c}_{i\sigma}+ h.c.),     
\label{eq:h4}
\end{eqnarray}  
where $ t_{ik} $s are hopping integrals where we consider them in the wide-band limit.

The forth term represents the free Hamiltonian of the single mode phonon bath and the fifth term is the electron-phonon interaction Hamiltonian, that is,
\begin{eqnarray}
\hat{H}_{ph}=\Omega \left( \hat{b}^{\dag} \hat{b}+\frac{1}{2}\right) ,     
\label{eq:h5}
\end{eqnarray} 
and
\begin{eqnarray}
\hat{H}_{e-ph}=\sum_{i,j,\sigma } \gamma \hat{c}_{i\sigma}^{\dag} \hat{c}_{j\sigma} \left( \hat{b}^{\dag}+\hat{b}\right) ,     
\label{eq:h6}
\end{eqnarray} 
where $ \hat{b} $ ($\hat{b}^{\dag}$) is the annihilation (creation) operator of phonons, $ \Omega $ is phonon frequency and $ \gamma $ is the strength of electron-phonon coupling.

Finally, the last term in the Hamiltonian represents the interaction of external classical electromagnetic fields with the QD in the electric-dipole approximation. In terms of creation and annihilation operators of the QD, it is given by
\begin{eqnarray}
\hat{H}_{e-l}=\sum_{\sigma}- \overrightarrow{\mu}.\overrightarrow{E}(t) (\hat{c}_{1\sigma}^{\dag} \hat{c}_{2\sigma}+h.c.),
\label{eq:h7}
\end{eqnarray}  
where $ \overrightarrow{\mu} $ is the electric-dipole matrix element between the two orbitals of QD and $ \overrightarrow{E}(t) $ is the external electric field.

The bias voltage, $ V $, is applied symmetrically to the right and left leads, thus their Fermi energies are, respectively, $\mu_{L}=eV/2$ and $\mu_{R}=-eV/2$. Moreover, the system is considered at zero temperature, so that the Fermi distribution of the leads are Heaviside theta functions ($ \theta (\mu_{\alpha}-\omega),\quad \alpha=R,L $), and we choose the system of units that $e=\hbar=1$.  

\subsection{The HF Approximation}
In order to consider the electron-phonon interaction and determine the optical properties of our model, we use the Keldysh formalism of non-equilibrium (contour-ordered) Greens functions (NEGF) which is an extremely useful  method for studying the non-equilibrium properties of many-body systems.

In our model, we deal with both electron and phonon GFs. The free Phonon GF on the Keldysh time contour is defined as
\begin{eqnarray}
D_{0c}(\tau,\tau^{'})=-i \left\langle  T_{c} \left\lbrace \hat{b}(\tau)+\hat{b}^{\dag}(\tau)\right\rbrace \left\lbrace \hat{b}(\tau^{'})+\hat{b}^{\dag}(\tau^{'})\right\rbrace \right\rangle~~
\label{eq:h8}
\end{eqnarray} 
It is a straight forward calculation to obtain the retarded, advanced and lesser GFs as functions of frequency (the greater GF is not independent of these tree):
\begin{eqnarray}
D_{0}^{r}(\omega)=\left[ D_{0}^{a}(\omega)\right]^{*}=\frac{1}{\omega-\Omega+ i 0^{+}}-\frac{1}{\omega+\Omega+ i 0^{+}}, 
\label{h9}
\end{eqnarray}
and
\begin{eqnarray}
D_{0}^{<}(\omega)=-2 \pi i \left[ n_{ph} \delta \left(\omega-\Omega\right) + \left( 1+n_{ph}\right) \delta \left(\omega+\Omega\right)\right],~~~~ 
\label{h10}
\end{eqnarray}
where $ n_{ph} $ is the expectation value of phonon number in free phonon bath, which is identically zero at zero temperature.

The electron GF on the Keldysh time contour is defined as
\begin{eqnarray}
G_{c,ij}(\tau,\tau^{'})=-i \left\langle  T_{c} \left( \hat{c}_{H,i\sigma}(\tau)\hat{c}_{H,j\sigma}^{\dag}(\tau^{'})\right) \right\rangle,~ i,j=1,2~~~~~~ 
\label{eq:h11}
\end{eqnarray}
where $ \hat{c}_{H,i\sigma}(\tau) $ ($ \hat{c}^{\dag}_{H,i\sigma}(\tau) $) is the annihilation (creation) operator in the Heisenberg representation on the Keldysh time contour. In our model the electron GFs are spin independent, so we show them by two by two matrices and drop their spin indices.

The non-interacting electron GFs, that is, the GFs in the absence of electron-phonon and electron-light interactions, are obtained by doing a straight forward calculation to be
\begin{eqnarray}
&&\hat{G}^{0r}(\omega)=\left[(\omega+i 0^{+}) \hat{I}-\hat{h}_{dot}-\hat{\Sigma}_{lead}^{r}\right]^{-1},
\\
&&\hat{G}^{0a}(\omega)=\left( \hat{G}^{0r}(\omega)\right)^{\dag},
\label{eq:h12}
\end{eqnarray}
and
\begin{eqnarray}
\hat{G}^{0<}(\omega)=\hat{G}^{0r}(\omega)\Sigma_{leads}^{<}(\omega)\hat{G}^{0a}(\omega),
\label{eq:h13}
\end{eqnarray} 
where $ \hat{I} $ is the unity matrix, $ \hat{h}_{dot} $ is the single particle non-interacting Hamiltonian of the QD, and the lead self-energies are
\begin{eqnarray}
\hat{\Sigma}_{leads}^{r}=-\frac{i}{2} \hat{\Gamma} ,
\label{eq:h14}
\end{eqnarray}
and
\begin{eqnarray}
\hat{\Sigma}_{leads}^{<}(\omega)=\frac{i}{2} \hat{\Gamma} \sum_{\alpha\in\lbrace R,L \rbrace} \theta(\mu_{\alpha}-\omega) ,
\label{eq:h15}
\end{eqnarray}
in which, $ \hat{\Gamma} $ is a two by two matrix that determines the level broadening of QD due to coupling to leads and its elements are given by
\begin{eqnarray}
\Gamma_{ij}(\epsilon)=2 \pi \sum_{k} t_{ik} t_{kj}^{*} \delta(\epsilon-\epsilon_{k}),
\label{eq:h16}
\end{eqnarray}
which is independent of $ \epsilon $ in wide-band approximation.

In order to consider the effect of electron-phonon interaction in electron GFs, we use the self consistent Hartree-Fock approximation to find an appropriate self-energy,$ \hat{\Sigma}_{HF} $, and insert it into the Dyson equation, which on the Keldysh contour reads as
\begin{eqnarray}
G_{c}=G^{0}_{c}+G^{0}_{c}\hat{\Sigma}_{HF,c}G_{c} ,
\label{eq:h17}
\end{eqnarray}   
where contour integrations are implicitly understood. 
The Hartree-Fock self-energy is the sum of Hartree and Fock terms. The Hartree self-energy on the Keldysh contour is
\begin{eqnarray}
&&\Sigma_{ij}^{H}(\tau,\tau^{'})=-i \gamma^{2} \delta_{ij} \delta^{c}(\tau-\tau^{'}) \int_{c} d\tau_{1} D_{0}(\tau,\tau_{1}) \nonumber \\ 
&&\sum_{l} 2 G^{<}_{ll}(\tau_{1},\tau_{1})  ,
\label{eq:h18}
\end{eqnarray} 
where $ \delta^{c} $ stands for the contour Dirac delta function. Because of this delta function, the lesser Hartree self-energy vanishes. Using the Langreth rules for analytical continuation and Fourier transforming the results, the retarded Hartree self energy in frequency domain is obtained as
\begin{eqnarray}
\Sigma_{ij}^{H,r}=-i \gamma^{2} \delta_{ij} D_{0} (\omega =0)  \int_{-\infty}^{\infty} \frac{d\omega^{'}}{2 \pi}\sum_{l} 2 G^{<}_{ll}(\omega^{'})  ,
\label{eq:h19}
\end{eqnarray} 
which is frequency independent. The electron population in the i-th orbital of the QD is
\begin{eqnarray}
n_{i}=-i \int_{-\infty}^{\infty} \frac{d\omega^{'}}{2 \pi} 2 G^{<}_{ii}(\omega^{'})  ,
\label{eq:h20}
\end{eqnarray} 
where the factor 2 is for spin. Combining Eqs.\ref{h9},\ref{eq:h19} and \ref{eq:h20}, we arrive at
\begin{eqnarray}
\Sigma_{ij}^{H,r}=\frac{-2 \gamma^{2}}{\Omega} \delta_{ij} \sum_{l} n_{l} .
\label{eq:h21}
\end{eqnarray} 

The Fock self-energy on the Keldysh contour is
\begin{eqnarray}
\Sigma_{ij}^{F}(\tau,\tau^{'})=i \gamma^{2} D_{0}(\tau,\tau_{'}) G_{ij}(\tau,\tau_{'})  ,
\label{eq:h22}
\end{eqnarray} 
Using the Langreth rules and Fourier transformation, the lesser and retarded self-energies are obtained to be
\begin{eqnarray}
\Sigma_{ij}^{F,<}(\omega)=i \gamma^{2}\int_{-\infty}^{\infty} \frac{d\omega^{'}}{2 \pi}G^{<}_{ij}(\omega^{'}) D_{0}^{<}(\omega-\omega^{'})  ,
\label{eq:h23}
\end{eqnarray} 
and
\begin{eqnarray}
&&\Sigma_{ij}^{F,r}(\omega)=i \gamma^{2}\int_{-\infty}^{\infty} \frac{d\omega^{'}}{2 \pi} \lbrace G^{r}_{ij}(\omega^{'}) D_{0}^{<}(\omega-\omega^{'}) \nonumber \\
&&+G^{r}_{ij}(\omega^{'}) D_{0}^{r}(\omega-\omega^{'})+G^{<}_{ij}(\omega^{'}) D_{0}^{r}(\omega-\omega^{'})\rbrace.
\label{eq:h24}
\end{eqnarray} 
Using Eq.\ref{h10}, we arrive at
\begin{eqnarray}
\Sigma_{ij}^{F,<}(\omega)= \gamma^{2} G^{<}_{ij}(\omega+\Omega),
\label{eq:h25}
\end{eqnarray} 
and
\begin{eqnarray}
&&\Sigma_{ij}^{F,r}(\omega)=\gamma^{2} G^{r}_{ij}(\omega+\Omega)+\nonumber \\
&& i \gamma^{2}\int_{-\infty}^{\infty} \frac{d\omega^{'}}{2 \pi} \lbrace G^{r}_{ij}(\omega^{'})+G^{<}_{ij}(\omega^{'}\rbrace  D_{0}^{r}(\omega-\omega^{'}).
\label{eq:h26}
\end{eqnarray} 
Adding the Hartree and Fock terms, we obtain the HF self-energy and consequently, the electron GFs as
\begin{eqnarray}
&&\hat{G}^{r}(\omega)=\left[ (\hat{G}^{0r}(\omega))^{-1}-\hat{\Sigma}_{HF}^{r}\right] ^{-1},
\label{eq:h27}
\\
&&\hat{G}^{a}(\omega)=\hat{G}^{r}(\omega)^{\dag},
\label{eq:h28}
\end{eqnarray}
and 
\begin{eqnarray}
\hat{G}^{<}(\omega)=\hat{G}^{r}(\omega) \left[ \hat{\Sigma}_{leads}^{<}(\omega)+\hat{\Sigma}^{F,<}(\omega)\right]  \hat{G}^{a}(\omega).
\label{eq:h29}
\end{eqnarray}
In numerical calculations, we have to find the electron GFs, HF self-energies and electron populations self consistently.

After doing these self-consistent calculations and obtaining the electron GFs, the left to right current could be computed from the formula
\begin{eqnarray}
I=\frac{i e}{ \hbar} \int_{\mu_{R}}^{\mu_{L}} \frac{d\omega}{2 \pi} Tr\left[ \hat{\Gamma} \left(\hat{G}^{r}(\omega)-\hat{G}^{a}(\omega)\right) \right],
\label{eq:h30}
\end{eqnarray}
where the trace is taken over the orbital degrees of freedom of the QD and a factor of 2 is already taken into account for spin.

\section{linear response to classical light}

The polarization of QD is defined by
\begin{eqnarray}
\overrightarrow{P}(t)=\overrightarrow{\mu} \sum_{\sigma} \left\langle \hat{c}_{H,1\sigma}(t)\hat{c}_{H,2\sigma}^{\dag}(t)+\hat{c}_{H,2\sigma}(t)\hat{c}_{H,1\sigma}^{\dag}(t)\right\rangle . \qquad
\label{eq:d12}
\end{eqnarray} 
Using equal time anti-commutation properties of annihilation and creation operators, Eq.\ref{eq:d12} can be written as\cite{AJWhite}
\begin{eqnarray}
\overrightarrow{P}(t)=4 Im\left[ \overrightarrow{\mu} \tilde{G}_{12}^{<} (t,t)\right],
\label{eq:d13}
\end{eqnarray}
where $ \tilde{G}^{<}_{12}(t,t) $ is the lesser GF between the two orbitals of QD when $ \hat{H}_{e-l} $ is taken into account.

Since, we are interested in the OPA cross section, we determine $ \overrightarrow{P}(t) $ to first order in $ \overrightarrow{E}(t) $. The contour-ordered GF to first order in $ \hat{H}_{e-l} $ is
\begin{eqnarray}
\hat{\tilde{G}}^{(1)}(\tau,\tau^{'})=\oint d\tau_{1} \hat{G}(\tau,\tau_{1}) \hat{h}_{e-l}(\tau_{1}) \hat{G}(\tau_{1},\tau^{'}),
\label{eq:d14}
\end{eqnarray}
where all the $ \tau $s lie on the Keldysh time contour, $ \hat{G} $ represents the HF GFs of the QD, and $ \hat{h}_{e-l} $ is
\begin{eqnarray}
\hat{h}_{e-l}(\tau)=\begin{pmatrix}
{0 & -\overrightarrow{\mu}.\overrightarrow{E}(\tau)\nonumber\\-\overrightarrow{\mu}.\overrightarrow{E}(\tau) & 0}
\end{pmatrix}.
\label{eq:d15}
\end{eqnarray}

Using the Langreth rules and doing the Fourier transformations, the first order polarization for $ \overrightarrow{E} $ in the direction of $ \overrightarrow{\mu} $ could be written as
\begin{eqnarray}
P^{(1)}(\omega)=2 i \mu^{2} E(\omega) (\gamma(-\omega,\omega)-\gamma^{*}(\omega,-\omega)),
\label{eq:d16}
\end{eqnarray}
where $E(\omega)$ is the Fourier transform of the electric field and
\begin{eqnarray}
&&\gamma(-\omega,\omega)=\int \frac{d\omega^{'}}{2 \pi}[ G_{12}^{r}(\omega+\omega^{'})G_{12}^{<}(\omega^{'})+G_{11}^{r}(\omega+\omega^{'})\times\nonumber\\
&&G_{22}^{<}(\omega^{'})+G_{12}^{<}(\omega+\omega^{'})G_{12}^{a}(\omega^{'})+G_{11}^{<}(\omega+\omega^{'})G_{22}^{a}(\omega^{'})].\qquad \label{eq:d17}
\end{eqnarray}
In terms of $ \gamma(-\omega,\omega) $, the frequency dependent OPA cross section is
\begin{eqnarray}
\sigma(\omega)=\frac{8 \pi \omega \mu^{2} }{c} Re[\gamma(-\omega,\omega)-\gamma^{*}(\omega,-\omega)].
\label{eq:d18}
\end{eqnarray}
If we have emission in our system, this OPA takes negative values.

\begin{figure}[ht!]
\includegraphics[width=8.5cm]{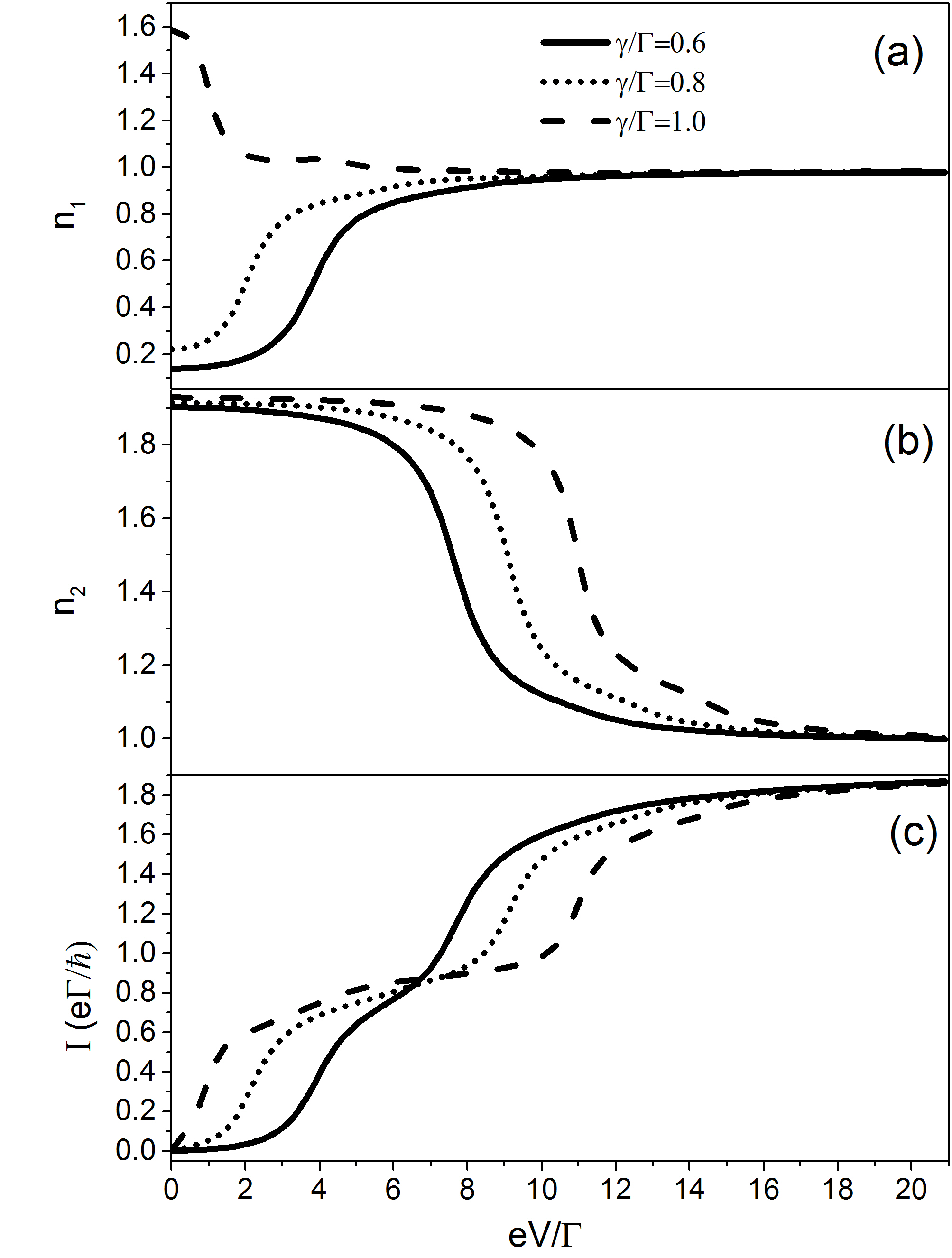}
\caption{\label{fig:one} (a), (b) and (c) are, respectively, the electron populations of the two orbitals of QD and  the current for $ \delta/\Gamma=3.0 $, $\Omega/\Gamma=1.8$ , $\gamma/\Gamma=0.6,0.8$ and 1.0, as functions of the bias voltage.}
\end{figure}

\section{Numerical Results}
In this section we present our numerical results. For our numerical calculations, we set $ \epsilon_{1}=-\epsilon_{2}=\delta $ so that the level spacing is $ 2\delta $. Moreover, we take all $ \Gamma_{ij} $s to be equal to a same value of $ \Gamma $ which is our energy unit. In Figs.\ref{fig:one} a, b and c, we show, respectively, the electron populations of QD levels, $ n_{1} $ and $ n_{2} $, and left to right current $ I $, as functions of applied bias voltage for $ \delta/\Gamma=3.0 $, $\Omega/\Gamma=1.8$ , $\gamma/\Gamma=0.6,0.8$ and 1.0. As we see in Fig.\ref{fig:one}a, for small electron-phonon couplings and at low bias voltages, the upper orbital of QD is almost empty, that is, $ n_{1} $ is small. But by increasing the strength of coupling, the upper orbital gets more populated even at small bias voltages. This behavior could be understood by noticing that the Hartree self energy, Eq.\ref{eq:h21}, is of minus sign and acts effectively as a gate voltage that lowers the onsite energies of QD, and by increasing the electron-phonon coupling strength, its value increases. By increasing the bias voltage, the electron populations of the two orbitals of QD change, until both $ n_{1} $ and $ n_{2} $ approach 1, which means the QD gets half filled.

\begin{figure}[ht!]
\includegraphics[width=8.5cm]{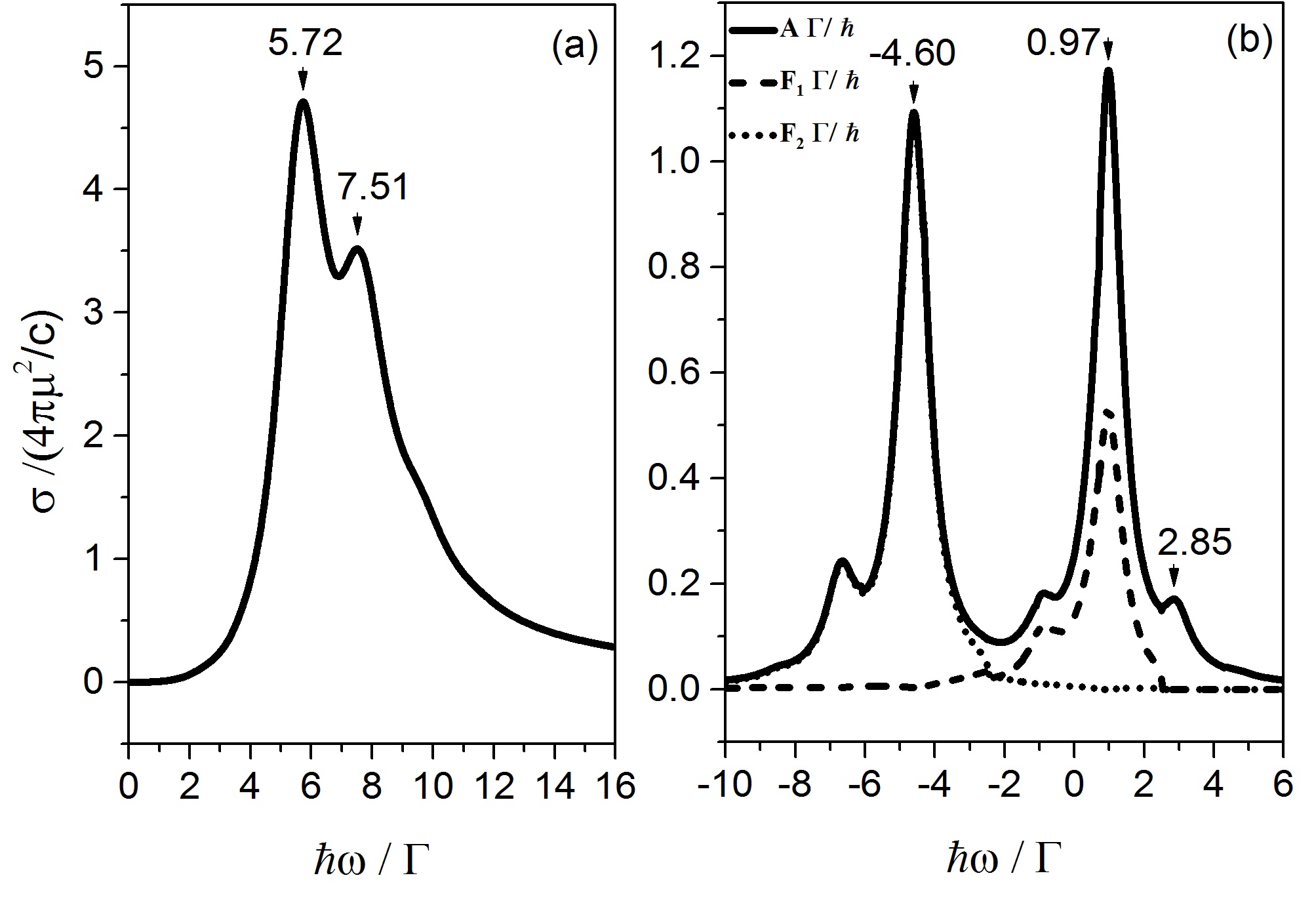}
\caption{\label{fig:two} (a)The OPA cross section as a function of frequency and (b) the density of states and filling functions  for $ \delta/\Gamma=3.0 $, $\Omega/\Gamma=1.8$ , $\gamma/\Gamma=0.8$ and $ eV/\Gamma=5.0 $. The arrows indicate the important peaks with their frequencies. The absorption peaks are explained by electron transitions between the states of QD. $ c $ is the speed of light.}
\end{figure}

At bias voltages where $ n_{1} $ is substantially greater than $ n_{2} $, the electrons could absorb light and make inter dot transition. In Fig.\ref{fig:two}a, we show this behavior by plotting OPA for $ \delta/\Gamma=3.0 $, $\Omega/\Gamma=1.8$ , $\gamma/\Gamma=0.8$ and $ eV/\Gamma=5.0 $. In Fig.\ref{fig:two}b, we show the density of states, $ A(\omega)=\frac{-2}{\pi} \sum_{i} Im(G^{r}_{ii}(\omega)) $, and in order to see what states are filled, we plot the filling functions $ F_{i}=\frac{-2i}{2 \pi} G^{<}_{ii}(\omega) $, whose integrals give the electron populations. Please note that the factor of 2 is considered for spin. Fig.\ref{fig:two}b reveals that the main peaks in OPA, Fig.\ref{fig:two}a, are caused by electron transitions between the states of QD. The two peaks of OPA at frequencies $ \omega/\Gamma=5.72 $ and 7.51, are due to electron transitions from state with energy $ \omega/\Gamma=-4.60 $ to states with energies $ \omega/\Gamma=0.97 $ and 2.85, respectively.

\begin{figure}[ht!]
\includegraphics[width=8.5cm]{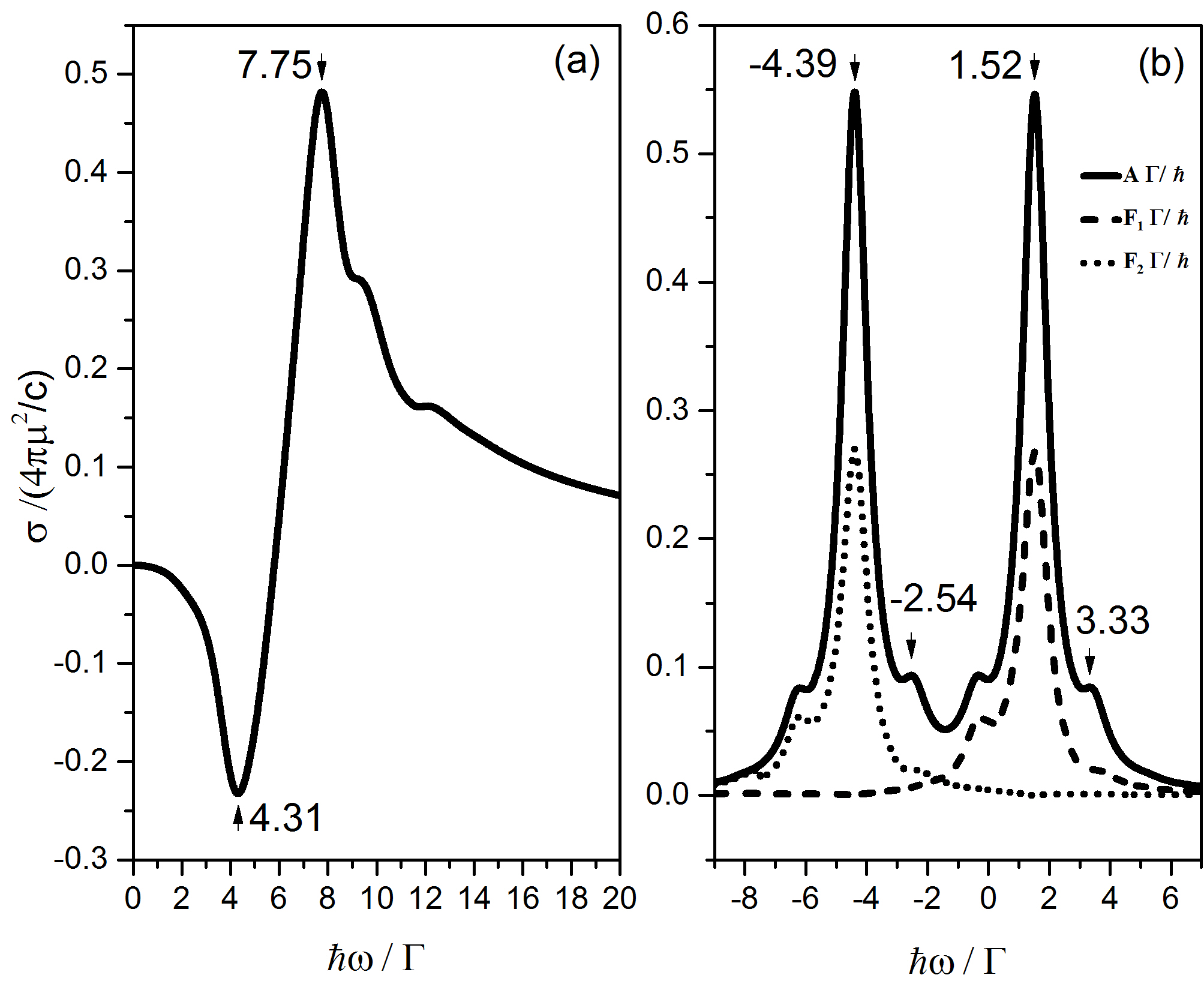}
\caption{\label{fig:three} (a)The OPA cross section as a function of frequency and (b) the density of states and filling functions  for $ \delta/\Gamma=3.0 $, $\Omega/\Gamma=1.8$ , $\gamma/\Gamma=0.8$ and $ eV/\Gamma=15.0 $. The arrows indicate the important peaks with their frequencies. The emission peak is explained by electron transmission from the higher main peak of density state to the side band at the right of the lower main peak. In addition, there is an absorption peak because electrons make transition from the lower main peak to the side band at right to the higher main peak. $ c $ is the speed of light.}
\end{figure}

At high bias voltages we have optical gain at some frequencies. In Fig.\ref{fig:three}a, we show OPA for $ \delta/\Gamma=3.0 $, $\Omega/\Gamma=1.8$ , $\gamma/\Gamma=0.8$ and $ eV/\Gamma=15.0 $. Even though $ n_{1} $ is smaller than $ n_{2} $, we have lasing with most strength at frequency $ \omega/\Gamma=4.31 $. Fig.\ref{fig:three}b shows the density of states and filling functions of QD. This figure reveals that the lasing takes place because one of the side-bands whose peak is at frequency $ \omega/\Gamma=-2.54 $ is almost empty and electrons could make transition from the higher orbital, whose main peak is at frequency $ \omega/\Gamma=1.52 $, to it. For the cases where level spacing is substantially more than the phonon frequency, since the side-bands are approximately at frequency distances of $ \Omega $ from the main peaks, we expect that the main optical gain occurs at frequency of about $ 2\delta-\Omega $, which is indeed the case. On the other hand, since the side-band right to the main higher peak of density of states is almost empty, electrons could absorb light and makes transition from lower main peak of density of states to this side-band. The distance between these two peaks is approximately $ 2\delta+\Omega $, which explains an absorption peak at almost this frequency.  

\begin{figure}[ht!]
\includegraphics[width=8.5cm]{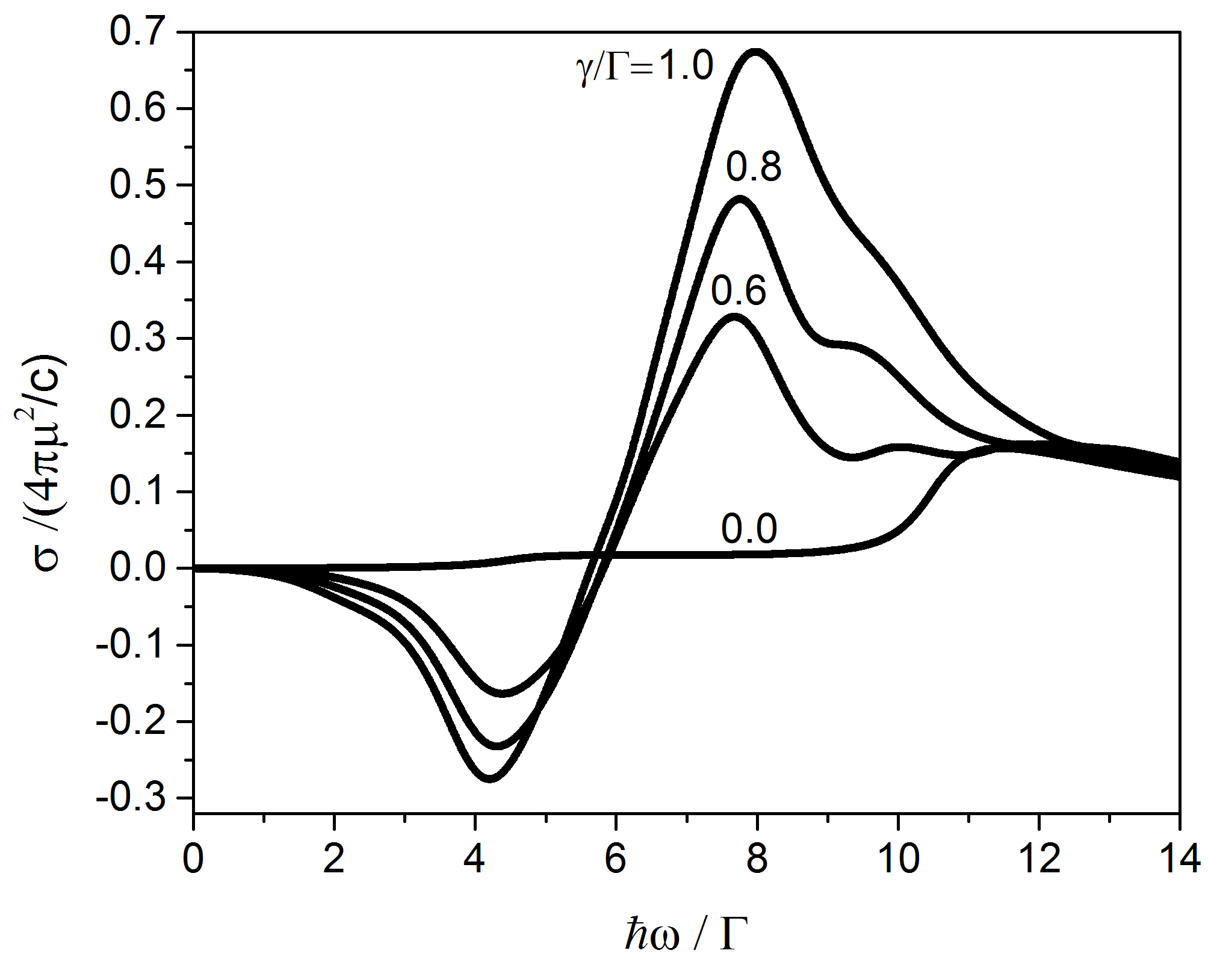}
\caption{\label{fig:four} OPA cross sections as functions of frequency for $ \delta/\Gamma=3.0 $, $\Omega/\Gamma=1.8$ ,$ eV/\Gamma=15.0 $, $\gamma/\Gamma=0.0,0.6,0.8$ and 1.0. The electron-phonon coupling strength is written on top of each curve. We see that the emission peak and the absorption resonance next to it, are completely due to electron-phonon interaction. $ c $ is the speed of light.}
\end{figure}

The emission and absorption peaks at approximately the frequencies of $ 2\delta\mp\Omega $ are completely due to electron-phonon coupling. In order to illustrate that, in Fig.\ref{fig:four}, we show OPA for $ \delta/\Gamma=3.0 $, $\Omega/\Gamma=1.8$ ,$ eV/\Gamma=15.0 $, $\gamma/\Gamma=0.0,0.6,0.8$ and 1.0. We see that increasing the electron-phonon coupling strength, the optical gain and the absorption peak next to it, both increase.

\section{Conclusions} 
In conclusion, we considered a two-level QD connected to two biased leads and coupled to a single-mode phonon bath. We obtained the I-V curve of our system and showed that depending on the used parameters, the populations of QD levels could be close to each other or differ substantially. By considering the linear response of the system to an incoming classical light, we concluded that at low bias voltages the QD absorbs light and we don't have optical gain. Moreover, at these low biases if the populations of the QD levels differ substantially, the light absorption is much stronger, that is, it would be easier for electrons to absorb light and make transition from the lower level to the higher one. On the other hand, at high enough voltages where both levels of the QD are almost half-filled, the system shows lasing. 

This lasing is seen because one of the lower side-bands is almost empty and electrons could make transition from the higher levels to it. On the other hand, one of the higher side-bands is also empty which explains an absorption peak. The frequency at which we have strong lasing is determined by the level spacing of QD, $ 2 \delta $, and the phonon frequency, $ \Omega $. For $ \Omega $ considerably less than $ 2 \delta $, this frequency is approximately $ 2 \delta-\Omega $, while we have an absorption peak at the approximate frequency of $ 2 \delta+\Omega $. Additionally, the strength of the lasing increases with increasing the electron-phonon coupling strength, provided that the QD remains almost half-filled.

\bibliographystyle{model1-num-names}
\bibliography{mp-006.bib}

\end{document}